\documentclass[9pt,twocolumn,twoside]{pnas-new}

\usepackage{amsmath}
\usepackage{aas_macros}
\usepackage{fontawesome}

\newcommand{\MT}{$M_\textup{200c}$}
\newcommand{\RT}{$R_\textup{200c}$}
\newcommand{\YT}{$Y_\textup{200c}$}
\newcommand{\cg}{$c_\textup{gas}$}

\newcommand{\be}{\begin{equation}}
\newcommand{\ee}{\end{equation}}
\def\Mpc{\, h^{-1} \, {\rm Mpc}}
\def\cg{c_\mathrm{gas}}

\def\Ms{\, h^{-1} \, M_\odot}

\templatetype{pnasresearcharticle} 
\setboolean{displaywatermark}{false}

\graphicspath{{./}{figures/}}

\title{Augmenting astrophysical scaling relations with machine learning: application to reducing the Sunyaev-Zeldovich flux-mass scatter}

\author[a,b,1]{Digvijay Wadekar}
\author[c]{Leander Thiele}
\author[d,e]{Francisco Villaescusa-Navarro}
\author[f,d]{J.~Colin Hill}
\author[e]{Miles Cranmer}
\author[d,e]{David N. Spergel}
\author[g]{Nicholas Battaglia}
\author[h,d]{Daniel Angl\'es-Alc\'azar}
\author[i]{Lars Hernquist}
\author[d,e,j]{Shirley Ho}

\affil[a]{School of Natural Sciences, Institute for Advanced Study, 1 Einstein Drive, Princeton, NJ 08540, USA}
\affil[b]{Center for Cosmology and Particle Physics, Department of Physics, New York University, New York, NY 10003}
\affil[c]{Department of Physics, Princeton University, Jadwin Hall, Princeton NJ 08544, USA}
\affil[d]{Center for Computational Astrophysics, Flatiron Institute, 162 5th Avenue, New York, NY, 10010, USA}
\affil[e]{Department of Astrophysical Sciences, Princeton University, Peyton Hall, Princeton NJ 08544-0010}
\affil[f]{Department of Physics, Columbia University, New York, New York 10027, USA}
\affil[g]{Department of Astronomy, Cornell University, Ithaca, NY 14853, USA}
\affil[h]{Department of Physics, University of Connecticut, 196 Auditorium Road, Storrs, CT, 06269, USA}
\affil[i]{Center for Astrophysics | Harvard \& Smithsonian, 60 Garden Street, Cambridge, MA 02138, USA}
\affil[j]{Department of Physics, Carnegie Mellon University, Pittsburgh, PA 15217}
\leadauthor{Wadekar}

\authordeclaration{The authors declare no conflict of interest. \\ \\Data deposition: The code and data associated with this paper is available at \url{https://github.com/JayWadekar/ScalingRelations_ML}.} 

\authorcontributions{D.Wadekar has led the project. D.Wadekar and L. Thiele performed the data analysis. J.C. Hill and D. Spergel suggested the research problem. F.Villaescusa-Navarro, D. Anglés-Alcázar and L. Hernquist provided the simulations. M. Cranmer and S.Ho suggested the machine learning algorithms. D.Wadekar contributed most to the manuscript writing. All authors commented on the manuscript.}

\correspondingauthor{\textsuperscript{1}Corresponding author. E-mail: jayw@ias.edu}


\significancestatement{
Two-dimensional power-law relationships discovered empirically in observed or simulated data are used for inferring properties of a wide variety of astrophysical objects (e.g., stars, supernovae and galaxies). More accurate relations, which are non-linear, or contain three or more variables, could easily have been overlooked, as they are difficult to find with manual data-analysis methods. We show that machine learning tools can expeditiously search for such relations in high-dimensional astrophysical data-spaces. In particular, we find improvements to previous relations which have been widely used for estimating masses of clusters of galaxies. Numerous upcoming observational surveys will target galaxy clusters, and our work enables their use to more accurately infer the fundamental properties of the Universe.}

\begin{abstract}
Complex astrophysical systems often exhibit low-scatter relations between observable properties (e.g., luminosity, velocity dispersion, oscillation period).  These scaling relations illuminate the underlying physics, and can provide observational tools for estimating masses and distances.
Machine learning can provide a fast and systematic way to search for new scaling relations (or for simple extensions to existing relations) in abstract high-dimensional parameter spaces. We use a machine learning tool called
symbolic regression (SR), which models patterns in a dataset in the form of analytic equations.
We focus on the Sunyaev-Zeldovich flux$-$cluster mass relation ($Y_\mathrm{SZ}-M$), the scatter in which affects inference of cosmological parameters from cluster abundance data.
Using SR on the data from the IllustrisTNG hydrodynamical simulation, we find a new proxy for cluster mass which combines $Y_\mathrm{SZ}$ and concentration of ionized gas ($c_\mathrm{gas}$): $M \propto Y_\mathrm{conc}^{3/5} \equiv Y_\mathrm{SZ}^{3/5} (1-A\, c_\mathrm{gas})$. $Y_\mathrm{conc}$ reduces the scatter in the predicted $M$ by $\sim 20-30$\% for large clusters ($M\gtrsim 10^{14}\, h^{-1} \, M_\odot$), as compared to using just $Y_\mathrm{SZ}$.
We show that the dependence on $c_\mathrm{gas}$ is linked to cores of clusters exhibiting larger scatter than their outskirts.
Finally, we test $Y_\mathrm{conc}$ on clusters from CAMELS simulations and show that $Y_\mathrm{conc}$ is robust against variations in cosmology, subgrid physics, and cosmic variance.
Our results and methodology can be useful for accurate multiwavelength cluster mass estimation from upcoming CMB and X-ray surveys like ACT, SO, eROSITA and CMB-S4. \href{https://github.com/JayWadekar/ScalingRelations_ML}{\faGithub}
\end{abstract}

\doi{\url{www.pnas.org/cgi/doi/10.1073/pnas.XXXXXXXXXX}}

\begin{document}

\maketitle
\thispagestyle{firststyle}
\ifthenelse{\boolean{shortarticle}}{\ifthenelse{\boolean{singlecolumn}}{\abscontentformatted}{\abscontent}}{}


\dropcap{A}strophysical scaling relations are simple low-scatter relationships (generally power laws) between properties of astrophysical systems which hold over a wide range of parameter values. Such relationships have a large number of applications:
$[i]$ inferring distances to objects, which is crucial for inferring cosmological parameters like the Hubble constant ($H_0$) (see e.g., the Leavitt period luminosity relation for Cepheids \citep{Lea12,Rei16,Rei19}, Phillips relation for supernovae \citep{Phi93}); $[ii]$ inferring properties of massive black holes (e.g., the black hole-bulge mass/velocity dispersion relation \cite{Kor13,Gre20,Hop07}); $[iii]$ inferring properties of galaxies (e.g., the Tully Fisher relation \citep{Tul77} and its baryonic analogue \citep{McG00} for spiral galaxies, the Faber Jackson relation \citep{Fab76}, the Kormendy relation or the more general fundamental plane relation \citep{Djo87,Dre87,Jor96,SheBer12} for ellipticals, the Color-Magnitude Relation); $[iv]$ providing insights into galaxy formation and evolution (e.g., the stellar to halo mass relation \cite{Wec18}); $[v]$ Inferring masses of galaxy clusters for cluster cosmology (e.g., the $Y-M$ relation \cite{Ade13,KraVik06,BatBon12}, $M_\mathrm{gas}-M$ relation \cite{Vik03,Voe04}, Mass-richness relation \cite{McC19}). Note that many of these relations have been discovered phenomenologically---often by trial and error---from observational data/simulations, rather than being derived from first principles \footnote{It is interesting to mention that, in some areas of physics, discovery of empirical relations has sometimes led to deep theoretical insights---take Kepler's laws giving inspiration to Newtonian mechanics, or the Planck equation (also an empirical function fit) aiding the development of Quantum Mechanics.}.

Most of the scaling relations found in astrophysics till now are power-law relations which involve only two variables. A reason for this could just be that it is easy to visually identify two-parameter relations in a dataset. 
There could exist many low-scatter relations with three or more variables in existing data which have been overlooked as it can be tedious to identify such relations with manual data analysis.
For instance, some of the popular two-parameter relationships were later shown to extend to three dimensions only by a more detailed subsequent analysis, e.g.,~the fundamental plane relationship for elliptical galaxies.
One of the traditional approaches to identify a high-dimensional non-linear hypersurface in a dataset is by looking at various 2D projection plots. This approach, however, becomes increasingly difficult and time consuming with larger datasets.

Machine learning (ML) tools can provide a faster and a more systematic approach to search for non-linear low-scatter relationships in abstract high-dimensional parameter spaces.
ML tools are increasingly useful as datasets available in astrophysics continue to grow in size due to advent of high-precision multi-wavelength surveys.
A particularly useful ML tool to search for new scaling relations, or to find extensions to existing ones, is symbolic regression (SR). SR identifies equations with parsimonious combinations of input parameters that have the smallest scatter with the given quantity of interest.


SR, also known as automated equation discovery, has been studied for decades in the context of scientific discovery, including early work creating the ``BACON'' algorithm~\citep{Langley1977BACONAP} and its later implementations including  COPER~\citep{kokar} and FAHRENHEIT/EF~\citep{LANGLEY1989283, Zembowicz}.
More recent work by \cite{bongard,Schmidt81} popularized SR for science, and introduced the software package Eureqa, which is a powerful (but proprietary) library still in use today.
This preceded significant interest from the ML community in advancing fundamental search techniques, including  \cite{eql, grammarvae,  brunton, koopman1,koopman2, deepymod,universalode, bayesianmachinescientist, 10.1162/evco_a_00278, Brunton3932, 2019arXiv190402107C, CraSan20, pysr, CraXu19,Vas22}. In parallel, these algorithms have been applied to a range of scientific problems, such as 
\cite{Del21, WadVil20b, astroexample2, astroexample3,  Sha21,LiuTeg11,Wil21, lemosRediscoveringNewtonGravity2022, butterBackFormulaLHC2021, gilpinChaosInterpretableBenchmark2021,CraSan20, cranmerDisentangledSparsityNetworks2021, cranmerHistogramPoolingOperators2021a, cravenDisentanglingDeepLearned2021, wernerInformedEquationLearning2021,Kro14}. It is worth mentioning that SR has been used in various astrophysical applications: modeling assembly bias \citep{WadVil20b, Del21}; estimating photometric redshifts of galaxies \cite{Kro14}; inferring universal subhalo properties \citep{Sha21}; modeling the concentration of dark matter from the mass distribution of nearby cosmic structures \citep{CraSan20}; discovering relationships in time-domain astronomy~\citep{astroexample2,astroexample3}; finding analytic forms of the one-point probability distribution function for neutrino-density fluctuations \citep{Ber21}; modeling the SFR density as a function of cosmological and astrophysical feedback parameters \citep{VilAngGen20}.

\begin{figure}
\centering
\includegraphics[scale=.7,keepaspectratio=true]{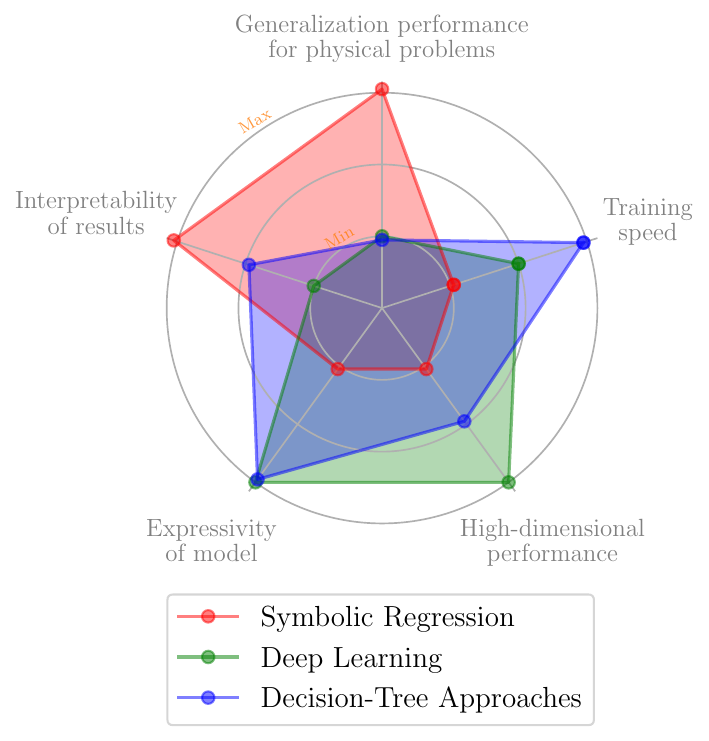}
\caption{Various aspects of the trade-offs between machine learning (ML) techniques. Symbolic regression can robustly be applied to datasets with only $\lesssim$ 10000 data points, each with $\lesssim$ 10 parameters. On the other hand, it can provide analytic equations that are readily interpretable and generalizable. We first use a decision-tree based approach called random forest regressor to narrow down the set of parameters that impact the scatter in the $Y$-$M$ relation. We then implement symbolic regression to find an analytic form for a cluster mass proxy using the pre-selected parameters.}
\label{fig:MLcomparison}
\end{figure}

In order to put SR in context, we illustrate tradeoffs in available ML tools along various dimensions in Fig.~\ref{fig:MLcomparison}. Deep learning tools like neural networks can handle very high dimensional inputs and large datasets, but are the least interpretable. SR lies on the opposite side of this spectrum: as of today, SR can be applied to datasets with only $\lesssim$ 10,000 data points, each with $\lesssim$ 10 parameters.
One must therefore simplify the problem or at times subsample the data in order to use SR on it. We follow the approach of  Ref.~\cite{WadVil20b}, where we first reduce the dimensionality of our dataset using a decision-tree based approach called a random forest regressor and then apply SR on it. Using the minimum set of relevant variables as input to SR is important to speed up its search for optimal equations.

We will focus on applying SR to find accurate expressions that relate properties of galaxy clusters to their masses. Galaxy clusters are the most massive bound structures in the Universe and their abundance as a function of mass is a very sensitive probe of cosmology \citep{Ade13,Ade15,Ade15b, Has13,Hil21,Boc15,Boc19,Pal20}. 
In the 2020s, many ongoing and upcoming
surveys (e.g., Rubin observatory, DES, HSC, DESI, ACT, eROSITA, SO, CMB-S4) will provide a wealth of multi-wavelength data on clusters.
If we can obtain robust mass estimates for these clusters from this data, we will be able to put very strong constraints on the nature of dark energy and neutrino masses \citep{Allen_2011,Sehgal_2011,Planck_2016,Bocquet_2019,Mad17,Mis18}. Cluster masses are typically inferred from properties easily measurable in observational surveys. For example, CMB surveys use the integrated electron pressure ($Y_\mathrm{SZ}$) via the mass-observable power-law relationship\footnote{In practice, the power-law exponent is calibrated with observational data, however, the actual fitted values are fairly close to 3/5, which is the prediction from virial theorem.}: $M_\textup{cluster}\propto Y^{3/5}_\mathrm{SZ}$ (the observable properties thus used are referred to as `mass proxies'). The scatter in these relationships affects the accuracy to which the masses---and thereby the cosmological parameters---can be inferred~\citep{Sha10} (e.g.,~the uncertainty in the scatter can be a source of systematic bias).
Therefore, an important property of a mass proxy is that the scatter in its relation with mass should be well-characterized and small.


 
A combination of observable properties (sometimes measured in different surveys) could sometimes provide a lower scatter mass proxy.  For example, X-ray studies show that the product of gas mass, $M_\mathrm{gas}$ and gas temperature, $T_X$, provides a lower scatter proxy than X-ray luminosity, gas mass or temperature: $Y_X\equiv M_\mathrm{gas} T_{X}$ \cite{KraVik06} \footnote{There have also been similar studies on augmenting the $Y-M$ relation \cite{Ver02,Afs08,Sha08,Yan10}; we will discuss them later in section~\ref{sec:Discussion}\ref{sec:63}.}. 
Recently, it has become possible to measure numerous properties of clusters: cluster electron pressure with SZ surveys, gas density and temperature profiles with X-ray surveys, density profiles with weak lensing surveys, spectra and color of galaxies in optical surveys, and diffuse synchrotron flux in radio surveys.
In order to construct an optimal mass proxy from these, one
encounters the following challenges: $(i)$ which particular properties in this large set to combine together? $(ii)$ what functional form should be used to fit the combination? 




ML methods can be useful for such problems. It is worth mentioning that there have been many recent ML motivated approaches to estimate cluster masses:
\citep{Gre19,CohBat20,Nta15,Nta19,Ho19,Ram20,CraSan20,Ram21,Gup20a,Gup20b,SuZha20,Yan20,Pab21,Nta22,Arm19,Fer22,deA22}.
Our goal in this paper is to model $M_\textup{cluster}$ by approximating the following function
\be
M_\textup{cluster} = f(Y^{3/5}_\mathrm{SZ}, \{i_\textup{obs}\})\, ,
\ee
with ML tools like random forests and symbolic regressors. $\{i_\textup{obs}\}$ is the set of various observable properties from multi-wavelength cluster surveys (e.g., gas mass, gas profile, richness, galaxy colors).
 As clusters are non-linear objects, there are no obvious first principles predictions for which properties in  $\{i_\textup{obs}\}$ should contribute. Furthermore, the high dimensionality of $\{i_\textup{obs}\}$ makes this a complex and challenging problem for traditional methods.


The paper is organized as follows. In Section~\ref{sec:ClusterData}, we briefly describe the cluster data that we use from various hydrodynamical simulations. In Section~\ref{sec:MassProxy}, we present an overview of mass proxies. We then discuss an overview of our ML techniques in Section~\ref{sec:ML} and show the results for cluster mass prediction in Section~\ref{sec:results}. We describe our reasoning behind using cluster concentration in Section~\ref{sec:Discussion}, and we conclude in Section~\ref{sec:Conclusions}.

\section{Cluster data and properties}
\label{sec:ClusterData}

In this section, we provide a brief description of the cluster data that we employ in our analysis.
We use the TNG300-1 simulation (hereafter TNG300) produced by the IllustrisTNG collaboration \citep{Nel19,Pil18,Spr18,Nel18,Nai18,Mar18,PilSprNel1801,Wei17}\footnote{IllustrisTNG: \url{https://www.tng-project.org/data/}}, which is run with the moving mesh AREPO code \citep{Spr10, Wei20}. We use the cluster samples from two different snapshots at redshifts $z=\{0\, ,\, 0.7 \}$ in our study. 

We also use clusters from the CAMELS suite of simulations \citep{VilAngGen20,Vil21}\footnote{CAMELS: \url{https://camels.readthedocs.io}}, which consists of more than 2,000 hydrodynamic simulations (each simulation box has length 25 $\Mpc$) run with different baryonic feedback and cosmological parameters, and with varying initial random seeds. CAMELS contain two distinct simulation suites, depending on the code used to solve the hydrodynamic equations and the subgrid model implemented: $(i)$ CAMELS-SIMBA, based on the GIZMO code \citep{Hop15,Hop17} employing the same sub-grid model as the flagship SIMBA simulation \citep{Dave2019};
$(ii)$ CAMELS-TNG, based on the AREPO code employing the same sub-grid model as the flagship IllustrisTNG simulations.
Let us provide one example to highlight the substantial differences in these models: feedback from active galactic nuclei (AGN) is implemented considering Bondi accretion and spherical symmetry in
IllustrisTNG \citep{Wei18}; while SIMBA implements gravitational torque accretion of cold gas and collimated outflows and jets from AGN \citep{Ang17}. We use clusters in the $z=0$ snapshots of the latin hypercube set for our analysis. (See \cite{VilAngGen20} for further details on the CAMELS simulations.)

For all the simulations, we work with halos identified by the FOF (friends-of-friends, also referred to as single linkage hierarchical clustering \cite{everitt2011cluster}) algorithm with linking length 0.2. We
choose the centers of clusters to be the locations of the minimum gravitational potential within the FOF volume. 
Note however that, to calculate properties of clusters mentioned later in this section, we do not use the FOF volume, but instead use use the spherical definition of clusters (we refer the reader to Ref.~\cite{Tin08} for the advantages of using a spherical halo definition over the FOF volume).
We use the boundary $R_\textup{200c}$ to define the cluster radii\footnote{\RT\ is the radius enclosing an overdensity $\Delta = 200$ with respect to the critical density of the Universe.}. $M_\textup{200c}$ is the mass of all the particles (dark matter, gas, stars and black holes) within $R_\textup{200c}$ of the center of the halo.
Note that we will use the 3D data of clusters in this paper; in reality, however, projected properties, instead of 3D, are measured in surveys; we will test our results for that case in a future study.
 We show the number of clusters as a function of their masses in SI Appendix Fig.~S1.  Let us now discuss the cluster properties we use in our study.\\

\noindent \textit{(i) Integrated electron pressure:}
CMB photons are scattered by high energy electrons 
in the plasma inside clusters due to inverse Compton scattering. This phenomenon is known as the thermal Sunyaev-Zeldovich (tSZ) effect and it induces a shift in the energy of the scattered CMB photons \citep{Sun70}.
Such a shift is typically parameterized by the integrated Compton-$y$ parameter ($Y_\mathrm{SZ}$) and can be directly measured in SZ surveys. We measure a 3D analogue of it in simulations, as given by 
\be
Y_{200c}= \frac{\sigma_\textup{T}}{m_e c^2}
\int_0^{R_{200c}} P_e (r)\, 4 \pi r^2 dr
\label{eq:Y200}
\ee
where $\sigma_\textup{T}$ is the Thomson cross section, $m_e$ is the electron mass, $P_e$ is the electron pressure and $c$ is the speed of light. Note that we use the group\_particles code\footnote{\url{https://github.com/leanderthiele/group_particles}} to obtain $P_e(r)$ (and most other properties mentioned in this section) from the simulation data.\\

\noindent \textit{(ii) Ionized gas mass:}
We calculate the cluster ionized gas mass ($M_\mathrm{gas}$) as
\be
M_\mathrm{gas}(r<R)= \frac{2}{1+X_H}\, m_p \int_0^{R} n_\textup{e} (r)\, 4 \pi r^2 dr
\label{eq:Mgas}
\ee
where $n_e$ is the free electron number density profile, $X_H=0.76$ is the primordial neutral hydrogen fraction, and $m_p$ is the proton mass.
Note that we derive $M_\mathrm{gas}$ from the electron density profile of a cluster 
in order to mimic the $M_\mathrm{gas}$ measurements from X-ray surveys (where $n_\textup{e} (r)$ is derived by de-projecting of X-ray surface brightness profiles \citep{Voe04,Cro06}).\\

\noindent \textit{(iii) Cluster concentration:}
We use different versions of the cluster concentration in this paper.
For the main results, we use concentration corresponding to the gas profile: $c_\mathrm{gas}\equiv M_\mathrm{gas}(r\, <\, R_{200c}/2)/M_\mathrm{gas}(r\, <\, R_{200c})$.
We also perform additional cross-checks using the concentration obtained by fitting an NFW profile to the halos. In particular, we use $c_\textup{NFW} \equiv R_\mathrm{vir}/R_\mathrm{scale}$ ($R_\mathrm{vir}$ is the virial radius and $R_\mathrm{scale}$ is the Klypin scale radius \citep{Kly11} corresponding to the largest subhalo in the halo) measurements by \cite{Gab21}, which were obtained by running the Rockstar code \citep{Beh13} on the TNG300 halos.\\

\noindent \textit{(iv) Stellar mass:}
We calculate $M_*$ by summing over of the masses of all the star particles within \RT. Note that this quantity represents thus the total stellar mass in the cluster, not the stellar mass of the central galaxy.\\

\noindent \textit{(v) Cluster triaxiality:}
We generally expect clusters to be triaxial since they are formed by accretion along filaments that can impose a tidal gravitational force upon the forming clusters.
We first calculate the moment of inertia tensor using
\be
T_{ij}\equiv \sum_\alpha m_\alpha (x_{i,\alpha} -\bar{x}_i) (x_{j,\alpha} -\bar{x}_j)
\ee
where $\bar{x}_i$ is the coordinate of the center-of-mass of the cluster and $m_\alpha$ is the particle mass (we only use the particles within \RT\ of the cluster center in our calculations). We calculate $T_{ij}$ in two different ways: first, using all particle types (gas+stars+DM+black holes); second, using only the gas particles. We then calculate the triaxiality of the cluster as $\lambda_1/\lambda_3$ where $\lambda_i$ are eigenvalues of $T_{ij}$ ordered as $\lambda_1<\lambda_2<\lambda_3$. We also check our results with a different definition of triaxiality: $(\lambda_1-\lambda_3)/2/(\lambda_1+\lambda_2+\lambda_3)$.\\

\noindent \textit{(vi) Cluster richness:}
The richness of a cluster is the number of galaxies associated with it. We select the galaxies using the threshold $M_\star>10^9\, h^{-1} M_\odot$ and by requiring the centers of the galaxies to be within \RT\ of the cluster center. At $z=0$, this threshold yields a number density of galaxies in the simulation sample of $\sim 0.02\, (h/\textup{Mpc})^{3}$. 

\section{Mass proxies}
\label{sec:MassProxy}

\begin{figure}
\centering
\includegraphics[scale=0.55,keepaspectratio=true]{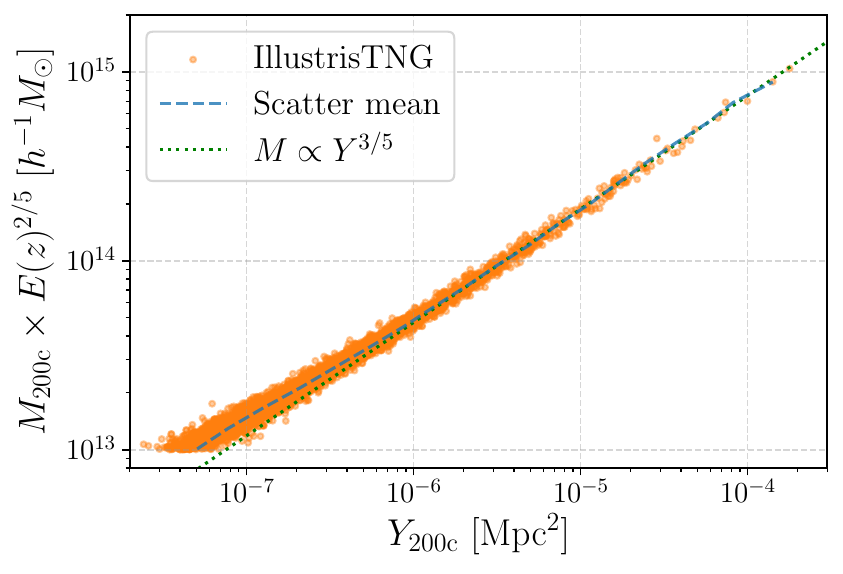}
\caption{$Y$-$M$ scaling relation in clusters from the TNG300 simulation at $z=\{0\, ,\, 0.7 \}$ (\YT\ [\MT] is the integrated Compton-$y$ parameter [cluster mass] within \RT). The self-similar power-law scaling relation normalized to the most massive halos is shown by the dotted green line. The goal of this paper is to improve this scaling relation in order to reduce its scatter and infer cluster masses more accurately.}
\label{fig:M_Y}
\end{figure}

Simple models of clusters based on the virial theorem (which assumes that the only source of energy input into the intra-cluster medium is gravitational) 
predict nearly self-similar relations between halo mass and various dynamic properties \citep{Kai86,KraBor12}. For example,
the scaling relation between cluster masses and temperature is given by \cite{Bry98}:
\be
T \propto (M\, E(z))^{2/3}
\label{eq:T}
\ee
where $E(z)\equiv H(z)/H_0=\sqrt{\Omega_m(1+z)^3+\Omega_\Lambda}$ for a flat Universe. Note that the temperature also depends on the value of $\Delta$ (the overdensity with respect to the critical density of the Universe used for defining the cluster boundary); we have absorbed this dependence under the proportionality sign. The scaling relation for the gas mass of a cluster is simply $M_\mathrm{gas} \propto M$. Using Eqs.~\ref{eq:Y200} and~\ref{eq:T}, one can write a scaling relation for the integrated Compton$-y$ parameter given by 
\be
Y_\mathrm{SZ} \propto M_\mathrm{gas} T \propto M^{5/3} E(z)^{2/3}
\ee
Scaling relations like these help in determining various possible proxies of cluster mass, e.g.,
\be
M\propto Y_\mathrm{SZ}^{3/5} E(z)^{-2/5}\, .
\label{eq:Y-M}
\ee

In addition to being motivated by idealized scaling relations, a mass proxy should have additional properties: $(i)$ \emph{Robustness}: it should be largely insensitive to limitations in our understanding of clusters, baryonic feedback effects, or their merger history, $(ii)$ \emph{Accuracy}: it should have a small and well-characterized scatter in the relation with mass, and $(iii)$ \emph{Low cost}: it should be observationally inexpensive in order to be applied for mass prediction of thousands of clusters.

$Y_\textup{SZ}$ satisfies all the aforementioned requirements. 
The self-similar evolution of the $Y_\textup{SZ}$-$M$ relation for clusters is also remarkably insensitive to baryonic physics like AGN feedback or radiative cooling (\cite{BatBon12,Sta10,Arn10,Fab11})).
The $Y_\mathrm{SZ}-M$ relation can be calibrated using two types of gravitational lensing measurements: CMB lensing measurements (which offer the advantage of a very well determined distance to the source plane)~\citep{Hu_2007,Baxter_2015,Geach_2017,Mad_2020}, and optical weak lensing surveys (which provide higher S/N measurements for individual clusters)~\citep{Hoekstra_2013,vdL_2014,Battaglia_2016,Med_2018,Schrabback_2018,Miyatake_2019}.
 Analogues of $Y_\textup{SZ}$ have therefore been used for cluster mass estimation in CMB surveys like Planck \citep{Ade13,Ade15,Ade15b}, ACT \citep{Has13,Hil21}, and SPT \citep{Boc15,Boc19}. It is worth mentioning that there are also proposals to self-calibrate the relation \cite{Maj03,Maj04}. An analogue of $Y_\textup{SZ}$ called $Y_X$ is also used in X-ray surveys for mass estimation \citep{KraVik06, Arn10}.
For a comprehensive review of the $Y_\mathrm{SZ}-M$ relation, see \cite{BatBon12}.  


We show the $Y_{200c}-M_{200c}$ relation from Eq.~\ref{eq:Y-M} for TNG300 clusters in Fig.~\ref{fig:M_Y} (see Eq.~\ref{eq:Y200} for the definition of $Y_{200c}$). For comparison, we also show the performance of other mass proxies like $M_\mathrm{gas}$ and cluster richness in
 SI Appendix, Fig.~S2.
 For a large region of parameter space in Fig.~\ref{fig:M_Y}, the clusters closely follow the self-similar scaling relation\footnote{A perceptive reader would notice that there is a deviation/break from the power law relation in Figure~\ref{fig:M_Y} for low mass clusters. This is because gas in the cluster gets ejected at low masses since the gravitational potential wells are comparatively shallower \citep{LovPil18,HilBax18,Le15,Gre15,Pan21}. We only focus on high mass clusters in this paper as only those are typically used in cosmological analyses; we have however modeled the deviations from self-similarity in a more recent paper \cite{WadThi22}.} with low scatter. Reducing the scatter further is imperative as the uncertainty in the mass-observable relation is currently the largest systematic uncertainty in cosmological analyses of galaxy clusters.

As we can see from Fig.~\ref{fig:M_Y}, $M_\textup{cluster}\propto Y^{3/5}_\textup{SZ}$ is a very good first approximation; we therefore train our ML models to approximate the following function based on the residuals:
\be
M_\textup{200c}/Y^{3/5}_\textup{200c} = g(\{i_\textup{obs}\})~.
\label{eq:intro}
\ee
In this way, we incorporate the domain knowledge (in our case the already well-established leading-order cluster physics) and use ML only to learn extensions to it.



\section{Machine learning techniques}
\label{sec:ML}

We now continue our discussion of machine learning (ML) techniques from the introduction section. In Fig.~\ref{fig:MLcomparison}, we had compared the ML techniques along two particular dimensions.
Deep neural networks (DNNs) are on one extreme: they can work with very high dimensional datasets or datasets with large sizes. There also have been many interesting applications of DNNs to cosmology (see e.g.,~\cite{HeLiFen1907,WadVil20,ZhaWanZha1902,GuiReyVil1910,Kri21,YipZha19,Kau21,ZamOkaVil1904,ModFenSel18,Kod20,TroFer19,Thiele_2020,CraSan20,cranmerDiscoveringSymbolicModels2020c,cranmerHistogramPoolingOperators2021a,thieleEquivariantModularDeepSets,LiNiCro20,BerSte19,Hor21,Paco_2021a,Paco_2021b,Lu_2021,Yin_2020a,Yueying_2021}). However, DNNs are notoriously difficult to interpret due to the high-dimensional parameter space of the model (typically $\gtrsim 10^6$ parameters). Furthermore, DNNs typically require very large datasets to train, whereas in our case, we only have $\sim$200 clusters with $M_{200c}>10^{14}\Ms$ in the TNG300 sample. 
We therefore used the two techniques detailed below, both of which can have better performance than DNNs on small datasets.

\subsection{Random forest}

A random forest regressor (RF) is a collection of decision trees; each tree is in itself a regression model and is trained on a different random subset of the training data \cite{Bre01_RF} (random forests can also be used for classification tasks, but here we use them for regression). The output from a RF is the mean of the predictions from the individual trees (a single decision tree is prone to overfitting and using the ensemble mean of different trees reduces overfitting) \cite{Eli08}.
RFs have been used for various applications in astrophysics: \cite{Mil15, Val19, Gre19,Aga18,LucPei18, MosNaa20, NadMao18,CohBat20,Muc20,Li22, Liu22}. As they allow one to easily infer the relative importance of each input feature, they are slightly better suited with regards to interpretability as compared to deep neural networks. Other advantages of decision tree based algorithms is that they comparatively much faster to train, and they do not require access to GPUs.

We use RF from the publicly available package \texttt{Scikit-Learn}\footnote{Random forest: \url{https://scikit-learn.org/stable/modules/generated/sklearn.ensemble.RandomForestRegressor.html}} \citep{Scikit}.
In order to check whether the results from the RF are robust to overfitting, we divide the data into two categories:
we use a sub-sample containing $\sim 40$\% of the clusters to train the RF, and the rest are used in testing the RF.
We show the results from the test set later in section~\ref{sec:results}\ref{sec:RFresults}. Note that we do not use RF for the final results of this paper, but only as a feature selection tool for making the application of symbolic regression easier.

\subsection{Symbolic regression}

Symbolic regression (SR) is a technique that approximates the relation between an input and an output through analytic mathematical formulae. The difference between using it versus ordinary ``least squares'' regression is that knowledge of the underlying functional form of the fitting function is not required a priori.
The advantage of using SR over other machine learning regression models 
is that it provides analytic expressions which can be readily generalized, and also facilitate the understanding of the underlying physics. One of the downsides of SR, however, is that the dimensionality of the input space needs to be relatively small. To overcome this, we first use the RF to obtain an indication of which parameters in the set of $\{i_h\}$ in Eq.~\ref{eq:intro} give the most accurate \MT. We then compress the $\{i_h\}$ set to include only the five most important parameters. Finally, we use SR on the compressed set to obtain an explicit functional form to approximate $f$ from Eq.~\ref{eq:intro}. We use the symbolic regressor based on genetic programming implemented in the publicly available \textsc{PySR} package\footnote{\label{PySR}\textsc{PySR}: \url{https://github.com/MilesCranmer/PySR}} \citep{pysr,CraSan20}.

Let us briefly describe the procedure to fit a function  with the \textsc{PySR} package.
First we specify the relevant input parameters (in our case, $\{ \cg,M_\mathrm{gas},M_*,c_\mathrm{NFW}\}$). We also need to specify unary and binary operators as input; we have chosen:
binary operators= [sum ($+$), multiplication($\cdot$), division($/$), power], and
unary operators=[negative, exponential, absolute value]. Using genetic programming, the SR then generates multiple iterations of formulae (e.g., $2.7\cdot M^2_{*}+\exp(M_\mathrm{gas}/ \cg)$). The best equations are decided based on their complexity and the specified loss function (equations which are the simplest and simultaneously give the least loss are preferable).

We use an analogue of the L1 loss function, given by
\be
\mathrm{Loss}=\sum_{i\in \mathrm{clusters}}\, w_i\, |M_i^\textup{true}- M_i^\textup{predicted}|
\label{eq:Loss}
\ee
The reason for choosing the L1 loss instead of L2  (i.e., Loss~$\propto |\Delta M|^2$) is that it is  as it is more robust to cases when the scatter is large. In other words, it is less susceptible towards outliers (see also other robust loss functions like Huber loss). As the number of halos decreases with their mass, we use the weights $w_i = M_i^{1/2}$ to upweight the high-mass halos (the weights also help in accounting for increased scatter towards low masses). Our primary focus in this paper is on clusters with $M\gtrsim 10^{14}\Ms$ as lower mass clusters are not used for probing cosmology (the lower mass regime is relatively more affected by AGN/supernova feedback). We specifically focus on improving $Y-M$ relation for low mass regime in a more recent paper \cite{WadThi22}. As separation between most clusters is too large for them to affect each other's evolution, we assume that their mass residuals are independent in the loss function in Eq.~\ref{eq:Loss}.

The complexity penalty of equations from SR is determined by the number of operators, free constants and variables in them. We use the default setting of equal complexity of individual operators, constants, variables (one also has the option to specify different values of complexity penalty to different operators, e.g., sin can be set to have three times the penalty of $+$). Note that there are traditional criteria to evaluate complexity of different fitting functions, e.g., Bayesian Information Criterion (BIC) or Akaike Information Criterion (AIC). However, such criteria typically only penalize the number of free constants and do not take into account the number of operators or variables in the equations, making them difficult to apply directly to output equations from SR.

It is worth mentioning that instead of needing to explicitly specify a parametric form like Eq.~\ref{eq:Loss} for the loss function, there are various non-parametric methods for fitting relations to data. A few examples are quantile regression, local regression models (e.g., Gaussian processes, local polynomial models like LOWESS)  \cite{kutner2005applied,sheather2009modern,fox2011r,Scikit}. Such methods are relatively advantageous to use when errors are heteroscedastic (i.e., the scatter is non-uniform, which is also the case for $Y-M$ relation at low masses), or the data contains outliers. These methods have been used in various astrophysical applications, e.g., \cite{Kaw17, Bra18, Li19}. However, we do not use them in our work as current SR packages require a parametric form of loss function to be specified (to our best knowledge, they are not currently designed to work with non-parametric loss functions).


\section{Results for $Y$-$M$ scatter}
\label{sec:results}

In this section, we compare the results from ML methods against the standard $Y$-$M$ relation. Most of the studies which carry out the analysis of $Y-M$ for cluster cosmology assume that the scatter is log-normal \cite{Hil21,Has13,Boc19} (see however \cite{Yan10,BatBon12}). We therefore choose to compare the performance of different mass estimation methods using the following statistic:
\be
\sigma_i\equiv \bigg[\frac{1}{N_i} \sum_{j}^{N_i}(\log M_j^\textup{true}- \log M_j^\textup{predicted})^2 \bigg]^{1/2}
\label{eq:Y_scatter}
\ee
where  $i$ corresponds to individual mass bins containing $N_i$ clusters (we used uniformly-spaced bins in log-space).
\subsection{Results from the random forest}
\label{sec:RFresults}
\begin{figure}
\centering
\includegraphics[scale=0.68,keepaspectratio=true]{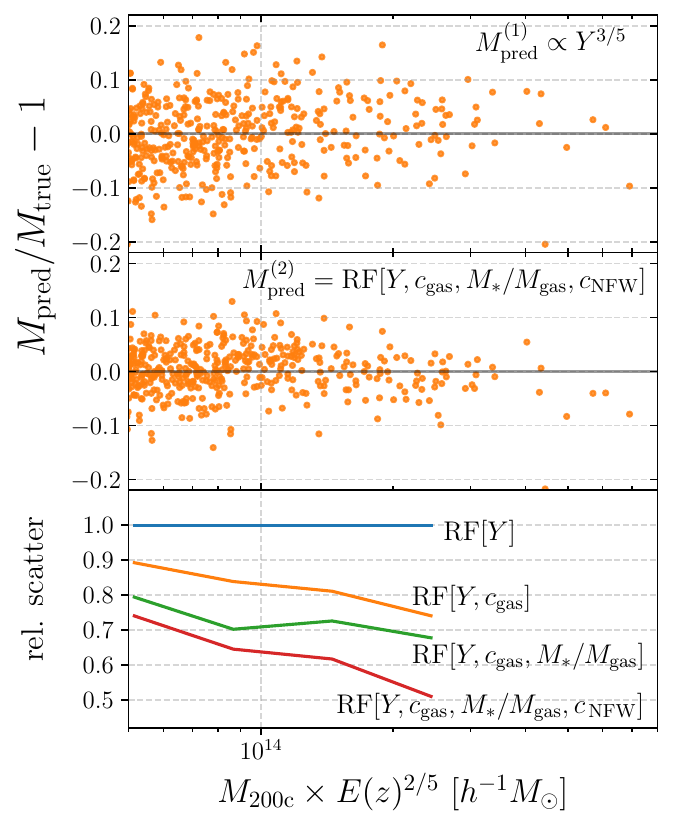}
\caption{Results from prediction of cluster masses with a random forest regressor (RF). Scatter in the predicted mass using the traditional Y-M relation (RF) is in the top (middle) panel. The bottom panel shows the effect of different sets of input parameters on the mass prediction. $\cg$ is the concentration of gas from Eq.~\ref{eq:GasConc}, $c_\mathrm{NFW}$ is the NFW concentration, $M_*$ ($M_\mathrm{gas}$) is the stellar (gas) mass within \RT. 
Overall, the RF improves mass prediction by $\gtrsim 30$\% as compared to the traditional scaling relation method.}
\label{fig:RF_scatter_reduction}
\end{figure}

We train the RF regressor using various cluster properties from Section~\ref{sec:ClusterData} and show results in Fig.~\ref{fig:RF_scatter_reduction}. In the bottom panel, we use Eq.~\ref{eq:Y_scatter} to calculate the scatter and show the relative improvement in the mass prediction (the improvement is $\gtrsim 30$\% for the best-case scenario). We do not compare the scatter for the very high-mass end as there are very few halos available to calculate the scatter robustly.

We also used cluster richness and triaxiality as input to the RF but did not notice any improvement in our results; 
we therefore do not show lines corresponding to them in Fig.~\ref{fig:RF_scatter_reduction}. We show the feature importance assigned by the RF to various input variables in SI Appendix Fig.~S3. We also tried using other galaxy properties (e.g., color of the brightest cluster galaxy), but we did not find any improvement in the scatter prediction.


\begin{figure}
\centering
\includegraphics[scale=0.6,keepaspectratio=true]{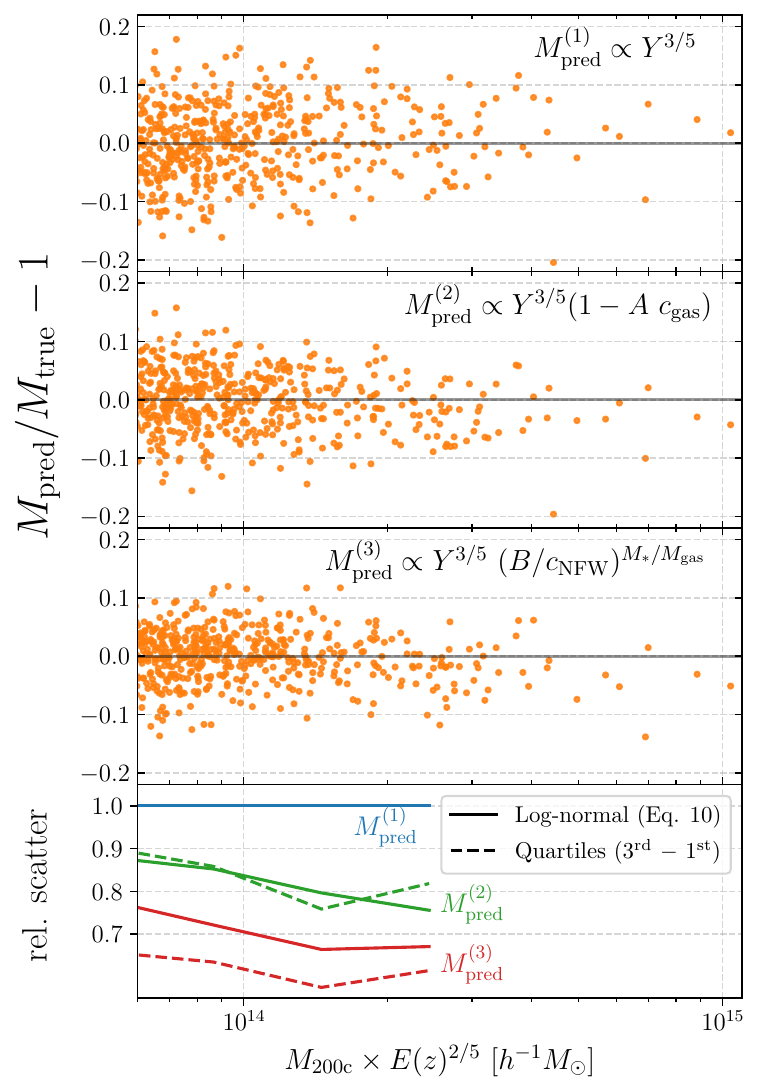}
\includegraphics[scale=0.6,keepaspectratio=true]{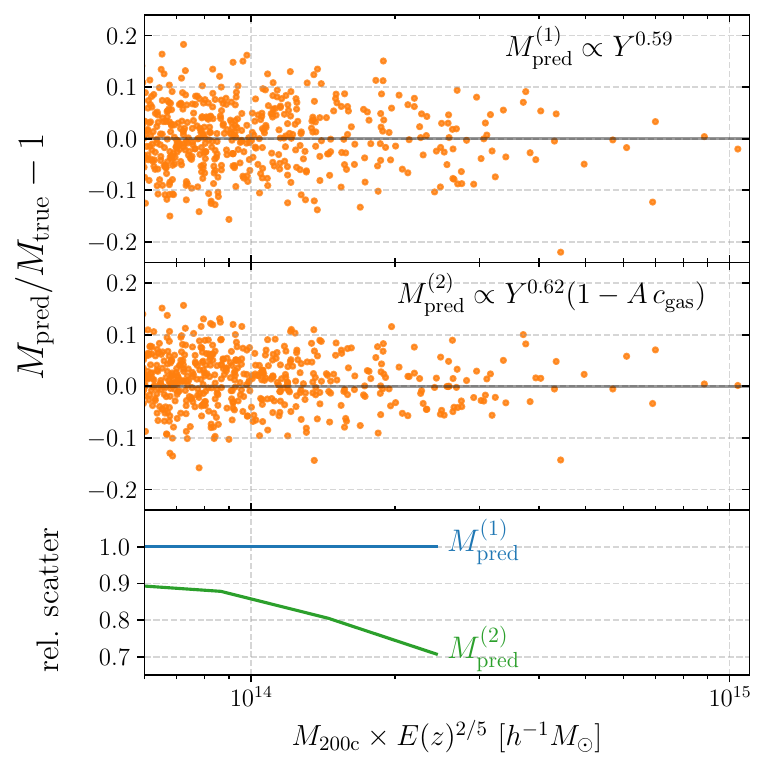}
\caption{\textbf{Top:} same as Figure~\ref{fig:RF_scatter_reduction}, but when the mass prediction is made using expressions from symbolic regression.
Second and third panels show our two best results from Eq.~\ref{eq:SyReg_conc} \& Eq.~\ref{eq:SyReg2} (additional results are shown in Fig.~S4 of the SI appendix). Fourth panel compares the scatter in the mass residuals (the scatter is calculated using two different methods).
We label the mass proxy in the second from top panel as $Y_\mathrm{conc}$. Introducting the term (1$-A\, \cg$) effectively down-weights the cluster cores in comparison to their outskirts (the cluster cores are relatively much noisier) and leads to a reduction in the scatter. \textbf{Bottom:} similar to the top case except the $Y-M$ power law slope is allowed to vary. Using $Y_\mathrm{conc}$ reduces the scatter by $\sim 25\%$ for \MT$\sim 2\times10^{14}M_\odot/h$.
}
\label{fig:SyReg_conc}
\end{figure}

\subsection{Symbolic regression}
Using the RF, we identified that the parameters: $c_\mathrm{gas}$, $M_*/M_\mathrm{gas}$ and $c_\mathrm{NFW}$ have the largest effect on the mass prediction. 
We now train the symbolic regressor to model the function in Eq.~\ref{eq:intro} using these properties
and obtain the results shown in Fig.~\ref{fig:SyReg_conc}.
Our main result of the paper is the following mass proxy which improves the cluster mass prediction as compared to using the standard $Y$-$M$ relation:
\be
M \propto Y_\mathrm{conc}^{3/5} \equiv Y_{200c}^{3/5}\, \left[1 - A\,c_\textup{gas}\right]
\label{eq:SyReg_conc}
\ee
where $c_\textup{gas}$ is related to the concentration of the halo gas profile and is given by
\be
c_\mathrm{gas}\equiv \frac{M_\mathrm{gas}(r\, <\, R_{200c}/2)}{M_\mathrm{gas}(r\, <\, R_{200c})}
\label{eq:GasConc}
\ee
where $M_\textup{gas}(r)$ is given by Eq.~\ref{eq:Mgas} and can be estimated from X-ray surveys. $A$ is a dimensionless parameter and we obtain the best-fit value $A=0.4$ for the TNG300 sample (we generally expect $A\in [0,1]$). We will discuss the physical explanation behind the better performance of $Y_\mathrm{conc}$ in Section~\ref{sec:Discussion}.

We also found that replacing $c_\mathrm{gas}$ in Eq.~\ref{eq:SyReg_conc} by an analogous parameter:
\be
c_{\,Y}\equiv \frac{Y(r\, <\, R_{200c}/2)}{Y(r\, <\, R_{200c})}
\label{eq:GasConc2}
\ee
gives a very similar improvement in the mass prediction. The advantage of using $c_{\, Y}$ over $c_\mathrm{gas}$ is that one does not need X-ray observations of clusters and SZ measurements alone are sufficient. On the other hand, it may not be straightforward to resolve scales of $R_{200c}/2$ (i.e., $\sim 0.7 R_{500c}$) in the observations of clusters from upcoming SZ surveys like SO and CMB-S4 due to their low resolution.\footnote{Looking further into the future, CMB-HD could provide high-resolution observations of clusters (in case full cluster pressure profile information is available, other ML tools like deep sets can be used to obtain even more accurate mass predictions).}

We also obtained the following mass proxy which has an even better performance than Eq.~\ref{eq:SyReg_conc}:
\be
M \propto Y_{200c}^{3/5}\, \left(\frac{B}{c_\mathrm{NFW}}\right)^ {M_*/ M_\mathrm{gas}}\, ,
 \label{eq:SyReg2}
\ee

where $B$ is another dimensionless constant (the best-fit value $B\sim50$ is used in the figure).
However, there are caveats regarding accurately estimating $M_*/ M_\mathrm{gas}$ or $c_\mathrm{NFW}$ from observational data. Analogues of $Y$ are typically estimated within $\lesssim 20$\% in current CMB surveys (see e.g., \cite{Hil21}). However, $M_*$ can only be estimated to within a factor of $\gtrsim$ 50\% accuracy with the current galaxy surveys (see e.g., \cite{Pal20,Hua18,Hua20,Hah22}). Therefore the mass estimation with Eq.~\ref{eq:SyReg2} could be dominated by observational uncertainties. More importantly, estimating the NFW concentration ($c_\mathrm{NFW}$) requires high-resolution lensing observations, and is therefore too expensive to measure for a large number of clusters.
Therefore, we will use $Y_\mathrm{conc}$ from Equation~\ref{eq:SyReg_conc} as our main result for the rest of the paper. 

In addition to using the lognormal assumption (Eq.~\ref{eq:Y_scatter}) to calculate the scatter in Fig.~\ref{fig:SyReg_conc}, we non-parametrically calculate the scatter using quartiles of the mass residuals and find a similar improvement when our new equations are used. We leave testing the assumption of lognormality of the $Y-M$ scatter to a future paper \cite{Rot23}.
Note also that we also obtained more complex equations as outputs from SR (some of them are shown in SI Appendix Fig.~S4). 
 However, given the large scatter already present in clusters from TNG300, the risk of overfitting goes up with increasing equation complexity. Hence, we show only the simplest expressions which have a relatively good performance. 

In cluster cosmology analyses, the power law index on $Y-M$ is usually not fixed to 3/5, but is fitted to data.
We therefore perform a test where we let the power law index vary. We use the \verb|scipy.fit| package and find the following best-fit relations: $M\propto Y^{0.59\pm0.002}$ and $M\propto Y^{0.618\pm0.002} (1 - [0.61\pm0.02]c_\mathrm{gas})$. Their performance is shown in the bottom panel of Fig.~\ref{fig:SyReg_conc}.

Due to the lack of clusters in the high-mass end of the TNG300 simulation, we are unable to compare the scatter between the different models. Cosmological simulations with a larger number of high-mass clusters (e.g., MillleniumTNG), or hydrodynamical zoom-in simulations centered on massive halos of a dark matter only simulation (e.g., the ones used in~\cite{Thiele_2020}) would be valuable to test our results. Generally, we expect results from machine learning algorithms to improve with a larger training dataset.

\begin{figure*}
\centering
\includegraphics[scale=0.7,keepaspectratio=true]{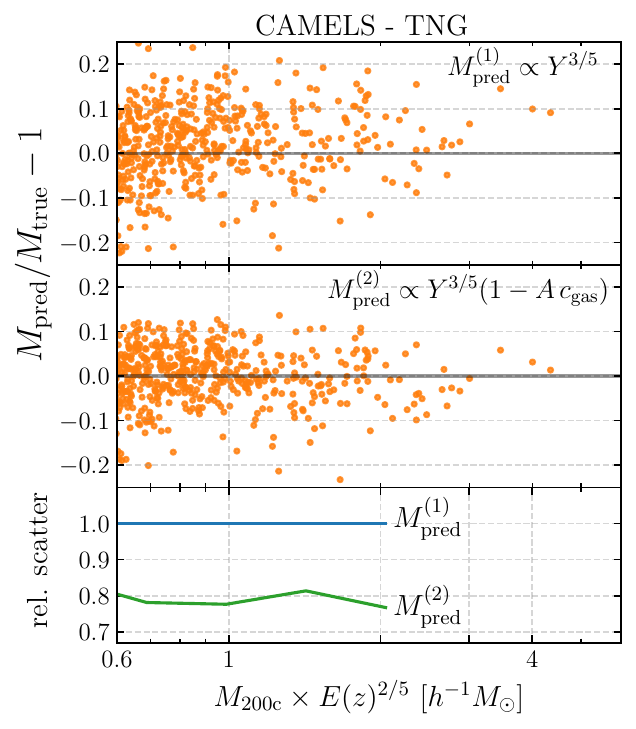}
\includegraphics[scale=0.7,keepaspectratio=true]{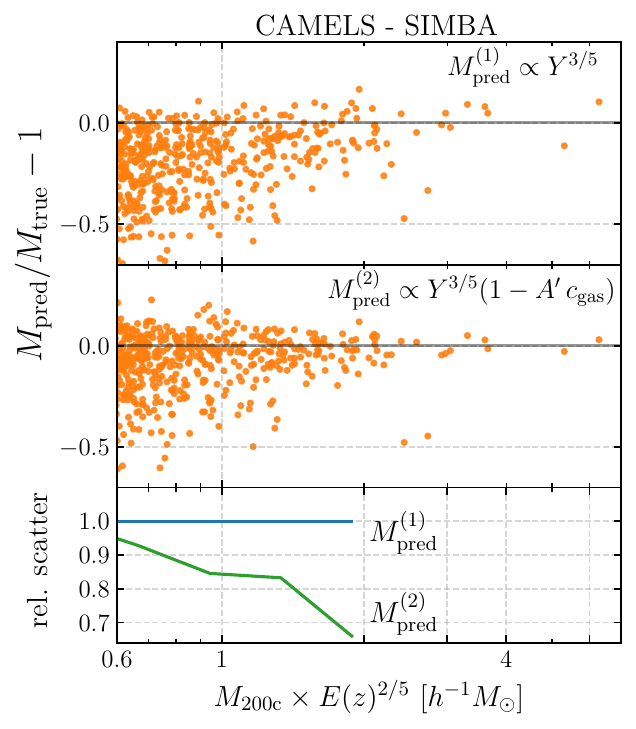}
\caption{Same as Figure~\ref{fig:SyReg_conc} but for halos in the CAMELS simulation suite instead of TNG300. As CAMELS includes variations in the baryonic feedback prescriptions in the hydrodynamic simulations, cosmological parameters and simulation initial seeds, the improvement upon using $M\propto Y_\mathrm{conc}$ is robust against these changes. Note also for CAMELS-SIMBA that $Y_\mathrm{conc}$ not only reduces the scatter, but also reduces the deviation from a power law for low \MT.}
\label{fig:camels}
\end{figure*}

\begin{figure}
\centering
\includegraphics[scale=0.7,keepaspectratio=true]{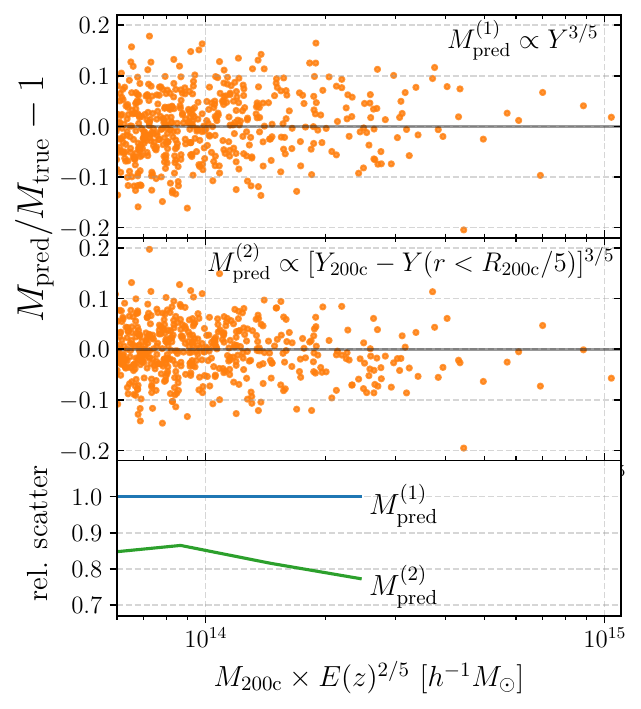}
\caption{Same as Figure~\ref{fig:SyReg_conc} but when the cores of the clusters are excised from the calculation of the integrated electron pressure. We see a roughly similar scatter reduction as in Figure~\ref{fig:SyReg_conc}. Directly excising the cores in upcoming CMB surveys is difficult because of their low resolution, hence using $Y_\mathrm{conc}$ is beneficial.}
\label{fig:CenterExcised}
\end{figure}

\begin{figure}
\centering
\includegraphics[scale=0.55,keepaspectratio=true]{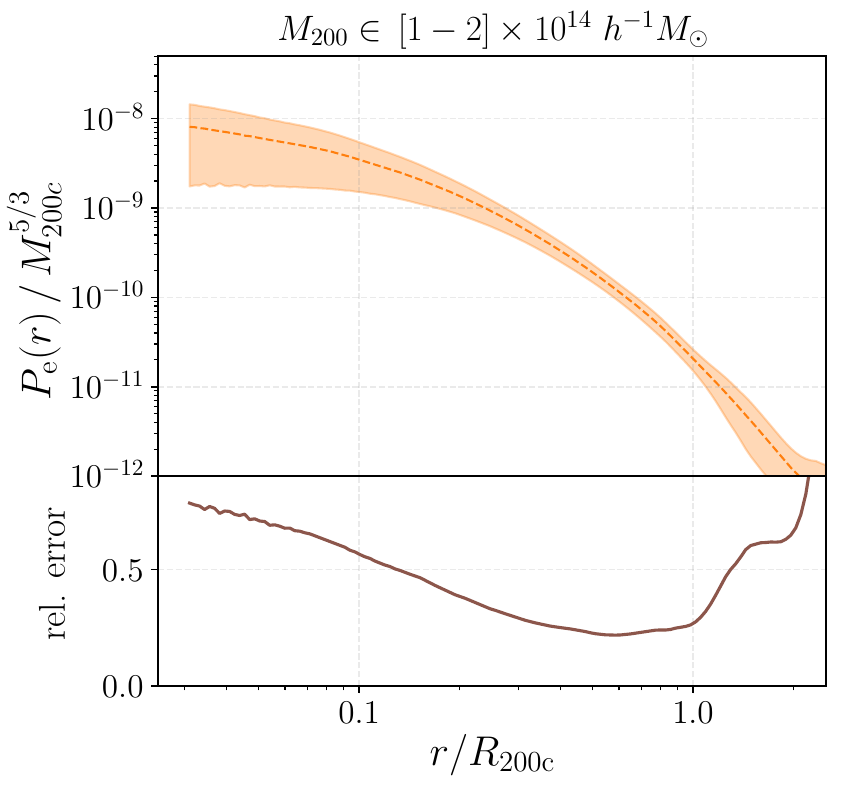}
\caption{Dependence of scatter with radius in the electron pressure profile ($P_e$) of clusters. We use the clusters from TNG300 in the specified mass range and show the mean and the 1$\sigma$ region of the pressure profile scaled by the cluster mass. The profiles have a large scatter in the innermost regions (cores), while the outer regions (until \RT) are relatively well equilibrated.}
\label{fig:Pprof_scatter}
\end{figure}

\subsection{Tests with CAMELS simulations}

Until this point, we showed results corresponding to the TNG300 simulation which uses a particular configuration of baryonic feedback parameters and a fixed cosmological model. However, the true nature of feedback in the Universe can be different, and we therefore want to test if the mass proxy $Y_\mathrm{conc}$ is robust to changes in feedback prescriptions. We therefore use the CAMELS suite of simulations which have varying cosmological and astrophysical feedback parameters, as well as varying initial conditions. We show our results for $z=0$ clusters in Fig.~\ref{fig:camels}. 

It is quite interesting that $Y_\textup{conc}$ consistently outperforms $Y_\text{SZ}$ even when the feedback prescriptions in the simulations are very different. Note that we did not retrain the symbolic regressor using the CAMELS dataset, we merely used Eq.~\ref{eq:SyReg_conc} and adjusted the constant $A$ to optimize our results. We found that using a larger constant $A'=0.8$ for CAMELS-SIMBA works better than using $A=0.4$ which was obtained for TNG300 (for CAMELS-TNG, however, the same constant: $A=0.4$ gives optimal results). This difference could be related to the scatter in the cores of SIMBA clusters being larger; we will return to this point in section~\ref{sec:Discussion}\ref{sec:61}. It is worth mentioning that the CAMELS simulations have a small box size (25 $\Mpc$) and there are very few high-mass clusters in the entire sample. It will be useful to check our results on the next iteration of the CAMELS simulations which will contain many more high-mass clusters.

\begin{figure*}
\centering
\includegraphics[scale=0.55,keepaspectratio=true]{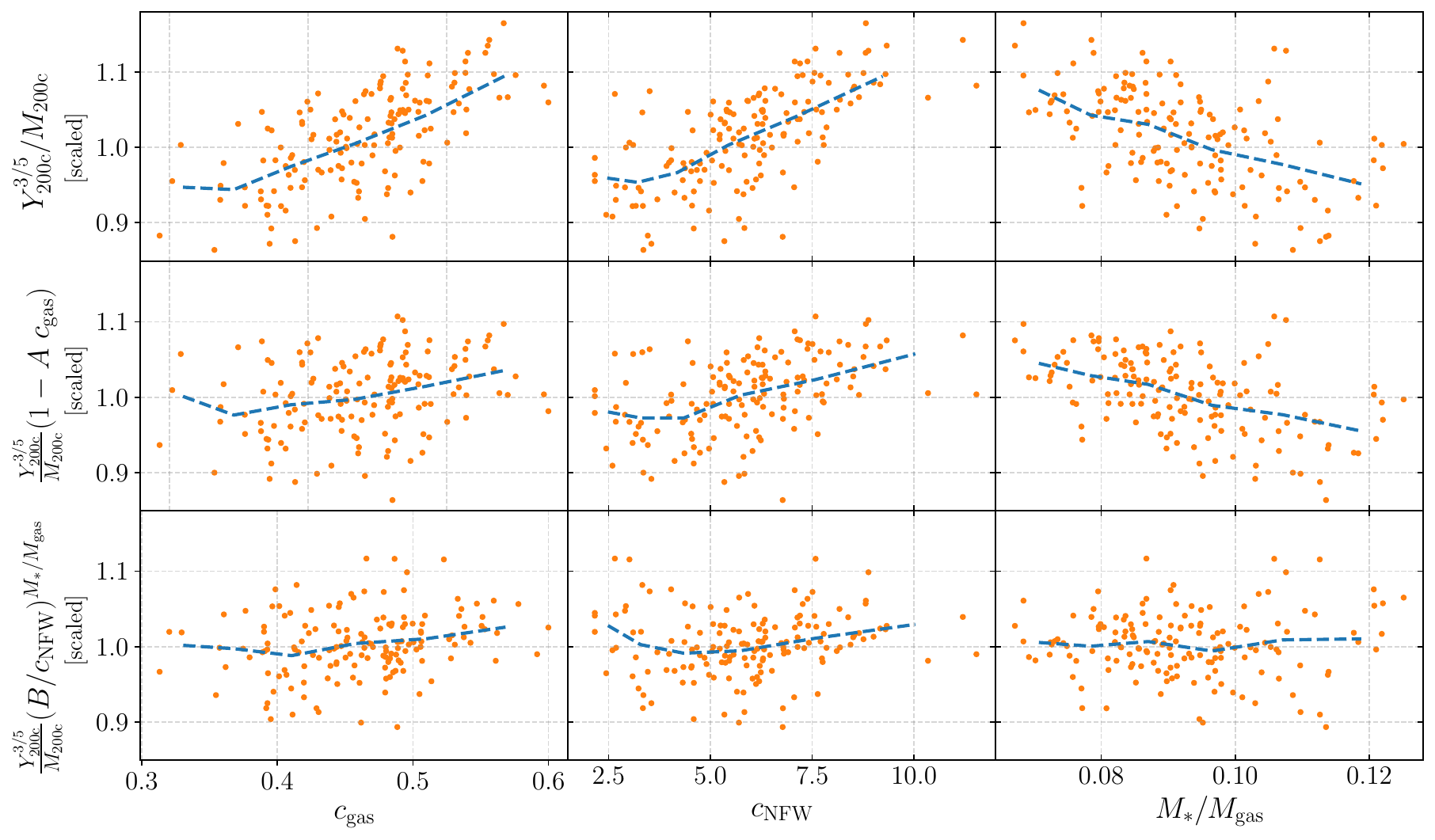}
\caption{\textbf{Top panel:} Dependence of the $Y$-$M$ relation on gas concentration ($c_\mathrm{gas}$), NFW concentration ($c_\mathrm{NFW}$) and stellar to gas mass ratio ($M_*/M_\mathrm{gas})$ for halos in the mass range $10^{14}\leq M \leq 2\times10^{14}\Ms$. The dashed lines show the mean of the scatter.
Higher $\cg$ (or $c_\mathrm{NFW}$) is related to increase in the density of the ionized gas and can be a result of more radiative cooling (which in turn increases \YT). Higher $M_*/M_\mathrm{gas}$ implies more gas being converted into stars, and is therefore associated with a decrease in \YT.
\textbf{Bottom panels:} Using $Y_\mathrm{conc}$ (middle) and Eq.~\ref{eq:SyReg2} (bottom) instead of $Y_{200c}$, for which the mean trends are comparatively weaker.
}
\label{fig:Y_conc}
\end{figure*}

\section{Discussion}
\label{sec:Discussion}

\subsection{Dependence on concentration}
\label{sec:61}
Having shown our results, let us now discuss some physical reasons behind the improvement in cluster mass prediction by taking into account concentration. For $A\in [0,1]$, the term (1$-A\, \cg$) contributes towards effectively down-weighting the cluster cores in comparison to their outskirts. Downweighting/excising the central regions is desirable because observed cluster profiles show a greater degree of similarity outside the core \citep{Vik06, Arn10,KraBor12,Ade13b}.
To verify this, we show in Figure~\ref{fig:CenterExcised} that the scatter in predicted mass is reduced when cluster cores are explicity excised from the calculation of \YT\ (Fig.~\ref{fig:CenterExcised} is for the TNG300 clusters, while the comparison with CAMELS clusters is shown in SI Appendix Fig.~S5).

Another way of verifying our results is to show the scatter in the pressure profile as a function of radius in the TNG300 clusters in Figure~\ref{fig:Pprof_scatter} (see also figure~4 of \cite{Ade13b} for comparison of pressure profile measurements from XMM-Newton and Planck).
Note that the cores are the regions of clusters which are the most sensitive to non-gravitational processes like radiative cooling and AGN feedback.  
Furthermore, simulations so far have not been able to convincingly reproduce the observed thermal structure of cool cores (see \cite{KraBor12}), and the observed scatter in 
cluster cores could be larger than that predicted in simulations \citep{Arn10}.
Given that $Y_\mathrm{conc}$ at least partly corrects for the cluster core effects, we expect it to perform better in case the scatter in cluster cores is larger. We also expect our method to work better in case $Y_{500c}$ is used instead of $Y_{200c}$ as the contribution from cluster cores is relatively larger for $Y_{500c}$.

We explicitly show the dependence of $Y$-$M$ relation on $c_\mathrm{gas}$, $c_\mathrm{NFW}$ and $M_*/M_\mathrm{gas}$ in the top panel of Figure~\ref{fig:Y_conc} for halos in the mass range $10^{14}\leq M \leq 2\times10^{14}\Ms$.
The bottom panel shows that $Y_\mathrm{conc}$ or Eq.~\ref{eq:SyReg2} takes into account a major part of these dependencies (which is responsible for the improvement in the cluster mass prediction due to them).

\subsection{Combining SZ and X-ray observations}
In the coming decade, numerous clusters will be probed with both X-ray (e.g., eROSITA survey \cite{Liu22, Chi22}) and SZ surveys (e.g., SO).
Let us now discuss ways in which these surveys can provide complementary information.
The advantage of X-ray surveys over SZ surveys is their higher resolution. On the other hand, their disadvantage is that they probe the cluster thermal energy indirectly (assumptions about the gas density and temperature profiles are needed to estimate the integrated pressure in X-ray surveys, whereas it is directly measured in SZ surveys).
Using $Y_\textup{conc}$ enables one to exploit this complementary behavior.

There are other advantages of combining SZ and X-ray surveys.
Cross-calibration across different wavelength measurements generally helps in minimizing the possible systematics in individual measurements such as projection effects (see e.g., \cite{Men10}).
Sometimes, $Y_\mathrm{spherical}$ reported by SZ surveys use an X-ray-derived estimate of the aperture size (as the cluster radii could be poorly measured by SZ surveys alone).
X-ray and SZ surveys have different redshift dependence: the selection function of SZ surveys flattens towards higher redshifts, while X-ray surveys favour low-redshift systems.
Combination of SZ and X-ray data can also help in removing outliers (e.g., recently merged clusters which deviate from the power-law relationship) and further tighten the $Y$-$M$ relation \citep{Yan10}. 

\subsection{Comparison with previous literature}
\label{sec:63}
Let us briefly mention some other proposals in the literature for augmenting the $Y$-$M$ relation. Refs.~\cite{Ver02, Afs08} proposed a fundamental plane relationship between $Y$, $M$ and the SZ half-light radius of the cluster.
\cite{Nel12} proposed augmenting the thermal pressure profile of clusters with a model for the non-thermal pressure in order to ameliorate the hydrostatic mass bias effect.
\cite{Sha08} noted that the NFW concentration can have an impact on the scatter in the $Y$-$M$ relation. \cite{Yan10} proposed augmenting $Y$-$M$ with a different form of cluster concentration: $R_{200}/R_{500}$. However, measuring this quantity requires high-resolution weak lensing data and this approach is therefore too expensive to be applied to a large number of clusters.
Our analysis provides a way of augmenting the $Y$-$M$ relation with properties that can be relatively easily measured in observational surveys. We also did a test with the random forest by adding analogues of the parameters proposed in the aforementioned studies for augmenting $Y-M$; we find that the RF predictions for cluster mass are improved only marginally (we show a comparison plot in SI Appendix Fig.~S6).

It is also worth mentioning that there have been studies augmenting other cluster scaling relations than $Y$-$M$, e.g.~\cite{Fuj18,Fuj18b,Fuj19} proposed a fundamental plane between cluster temperature, its mass and the scale radius of its matter profile. Recently, cluster NFW concentration was used in improving the model for the electron number density and pressure profiles of clusters \cite{Lee22}.

\section{Conclusions}
\label{sec:Conclusions}
Astrophysical scaling relations have a number of applications in inferring properties of stars, supernovae, black holes, galaxies and clusters. With the upcoming high-precision astronomical surveys, it is imperative to find ways to augment the existing scaling relations in order to make them more accurate. Machine learning can provide a fast and systematic approach to search for extensions to scaling relations in abstract high-dimensional parameter spaces.


We focused on searching for augmentations to the widely-used $Y_\mathrm{SZ}-M$ scaling relation in order to make mass prediction of galaxy clusters more accurate.
We first used a random forest regressor to search for a subset of parameters which give the most improvement in the cluster mass prediction (Fig.~\ref{fig:RF_scatter_reduction}).
We consequently used symbolic regression 
 and found a new mass proxy which combines $Y_\mathrm{200c}$ and gas concentration ($\cg$): $M \propto Y_\mathrm{conc}^{3/5}\equiv Y_\mathrm{200c}^{3/5} (1-A\, c_\mathrm{gas})$. 
$Y_\mathrm{conc}$ reduces the scatter in the mass prediction by $\sim 20-30$\% for large clusters ($M_{200c}\gtrsim 10^{14}\Ms$) at both high and low redshifts (Fig.~\ref{fig:SyReg_conc}). The new proxy exploits the complementary behavior of X-ray (high resolution but indirect probe of cluster thermal energy) and SZ (low resolution but direct probe of thermal energy) surveys.

We verified that $Y_\mathrm{conc}$ is robust against changes in both feedback parameters and subgrid physics by testing it with the CAMELS suite of simulations (Fig.~\ref{fig:camels}).
The dependence of $Y_\mathrm{conc}$ on $\cg$ is likely due to the cores of clusters being noisier (Fig.~\ref{fig:Pprof_scatter}), and we verify this explicitly by excising the 
cores of clusters (Fig.~\ref{fig:CenterExcised}). Our results and methodology can be useful for accurate multiwavelength cluster mass estimation from current and upcoming CMB and X-ray surveys like ACT, SO, eROSITA and CMB-S4.


\noindent \emph{Future work:} We use three dimensional cluster information (e.g., \YT) in this paper; but, in reality, projected properties of clusters (e.g., $Y_\mathrm{cylindrical}$) are measured in surveys; we will try to test our results for that case in a future study.
We focused on improving the $Y$-$M$ relation for high $M$ regime in this paper, but we use a similar ML motivated methodology for improving $Y$-$M$ in the low $M$ regime in a more recent paper \citep{WadThi22}. We could not robustly test $Y_\mathrm{conc}$ for very high mass clusters ($M \gtrsim 5\times 10^{14} \Ms$) due to lack of statistics, but we will do this test using clusters from the MilleniumTNG simulation (which has 15 times the volume of TNG300) in a separate upcoming paper \cite{Rot23}.

As cluster observations improve, we will be able to use ML techniques directly on observed quantities and find the lowest scatter relations between lensing masses, microwave and X-ray observables. Our methodology could also be useful for improving other widely-used astrophysical scaling relations for exoplanets, stars, supernovae, galaxies and clusters.


\acknow{First of all, we especially thank the anonymous referees for their critical comments as well as various useful suggestions on the manuscript. We also thank Daisuke Nagai, Nadia Zakamska, Matias Zaldarriaga, Tibor Rothschild, Joshua Speagle, Niayesh Afshordi, Suzanne Staggs, and Abhishek Maniyar for fruitful discussions.
DW gratefully acknowledges support from the Friends of the Institute for Advanced Study Membership. FVN acknowledges funding from the WFIRST program through NNG26PJ30C and NNN12AA01C.
NB acknowledges the support from NSF grant AST-1910021 and NASA grants
21-ADAP21-0114 and 21-ATP21-0129.
DAA was supported in part by NSF grants AST-2009687 and AST-2108944.
The work of SH is supported by Center for Computational Astrophysics of the Flatiron Institute in New York City. 
The Flatiron Institute is supported by the Simons Foundation. JCH acknowledges support from NSF grant AST-2108536.
We also thank Boryana Hadzhiyska, Will Coulton and Rachel Somerville for help with the ROCKSTAR catalogs corresponding to TNG halos.}

\showacknow{}

\bibliography{SZ}

\clearpage
\appendix
\setcounter{equation}{0}
 \setcounter{table}{0}
 \setcounter{figure}{0}

 \renewcommand{\theequation}{S\arabic{equation}}
 \renewcommand{\thefigure}{S\arabic{figure}}
 \renewcommand{\thetable}{S\arabic{table}}
 
 \begin{figure*}
\begin{center}
{\LARGE \bf Supporting information for `Augmenting astrophysical scaling relations with
machine learning'}
\end{center}
\end{figure*}
%

\begin{figure}
\centering
\includegraphics[scale=0.5,keepaspectratio=true]{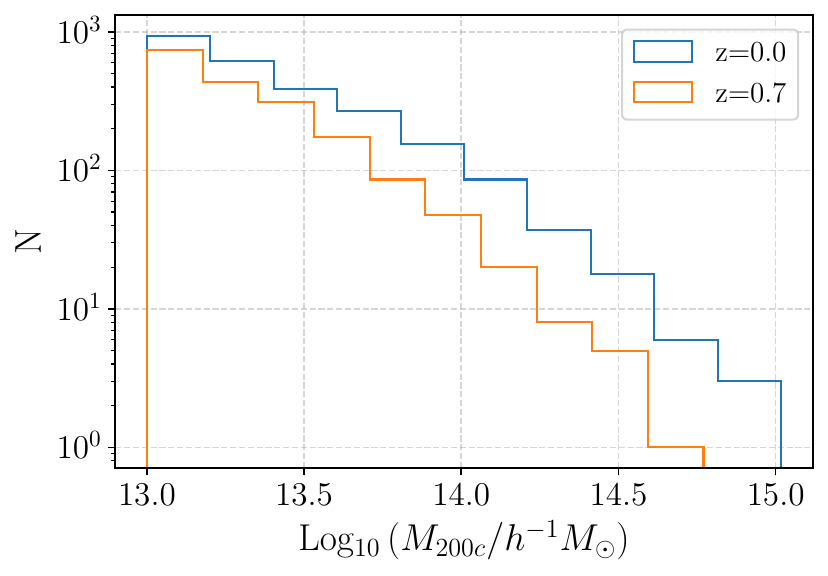}
\caption{The number of clusters as a function of their mass for the two IllustrisTNG300 snapshots which were used in the paper.} 
\label{fig:M_histogram}
\end{figure}

\begin{figure*}
\centering
\includegraphics[scale=0.4,keepaspectratio=true]{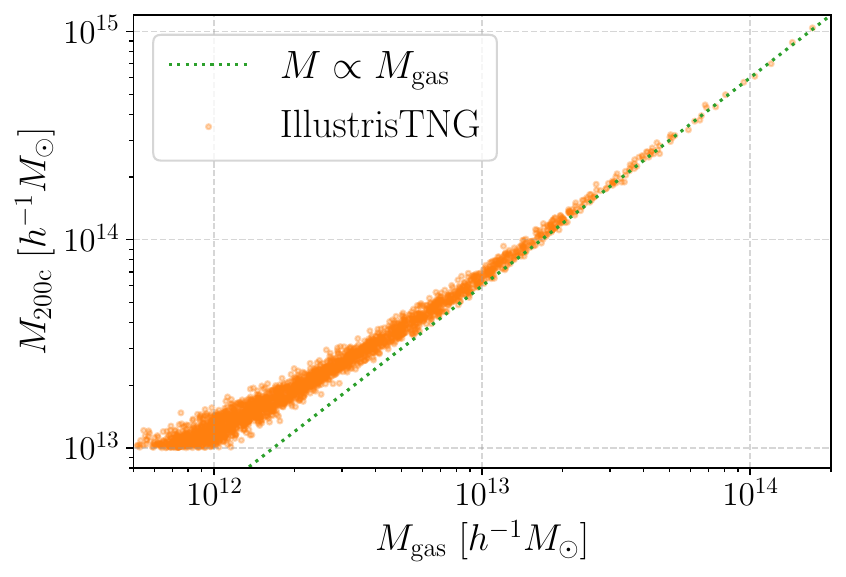}
\includegraphics[scale=0.4,keepaspectratio=true]{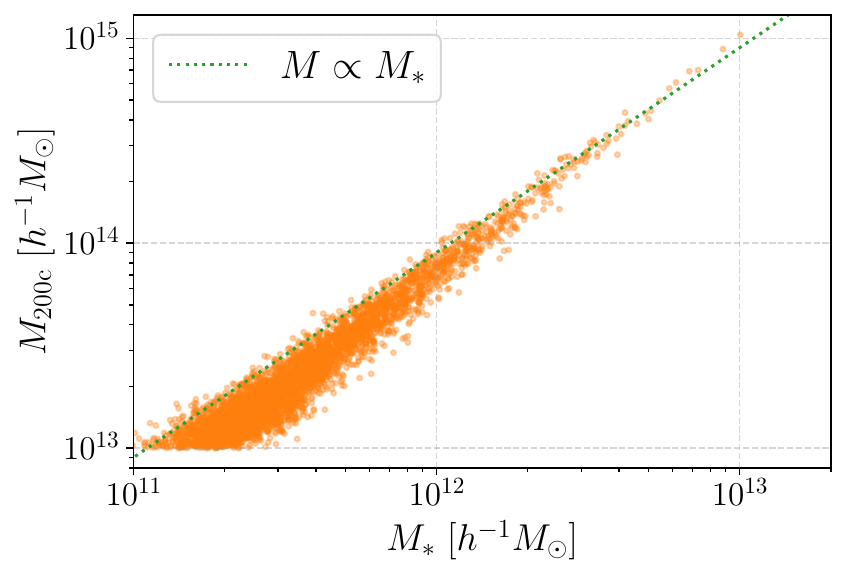}
\includegraphics[scale=0.4,keepaspectratio=true]{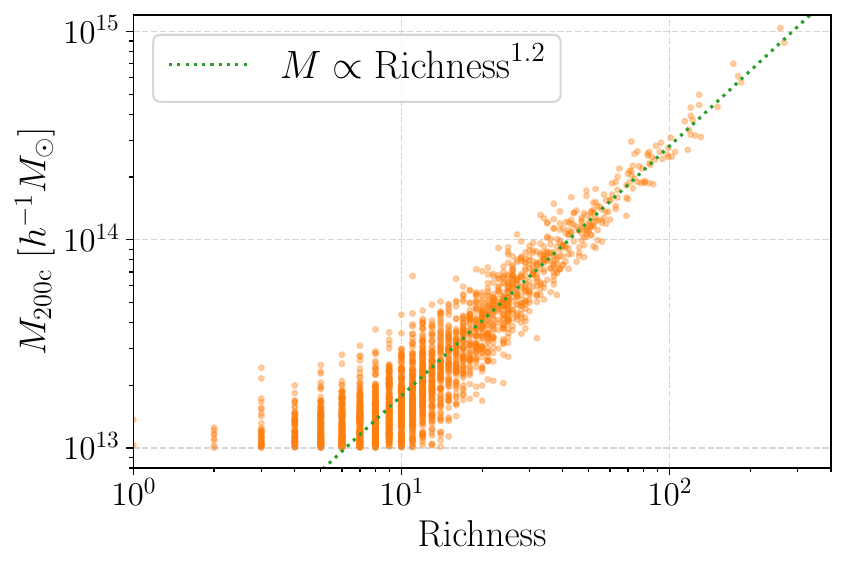}
\caption{Similar to Fig.~\ref{fig:M_Y} in the main text, but for scaling relations for other proxies of halo mass: $M_\mathrm{gas}$ from X-ray surveys, $M_*$ and richness (i.e., galaxy number counts) from galaxy surveys. The cluster data is from the TNG300 simulation at $z=0$. The power-law scaling relation normalized to the most massive halos is shown by the dotted green line.}
\label{fig:MassProxy}
\end{figure*}

In Fig.~\ref{fig:M_histogram}, we show the mass function of the clusters in two different snapshots of the IllustrisTNG simulation. We also show the relation of different observable properties of clusters to their mass in Fig.~\ref{fig:MassProxy}.

In Fig.~\ref{fig:M_Y} of the main text,
we showed the performance of \YT\ as a mass proxy. Fig.~\ref{fig:MassProxy} shows the comparison of other mass proxies. 
We see that the cluster richness has a much larger scatter than \YT, which makes the cluster mass estimation relatively less accurate. One additional issue in using cluster richness from galaxy photometric observations is that, because of the presence of background galaxies, it is not possible to state with absolute confidence that any given galaxy belongs to a given cluster.
$M_\mathrm{gas}$, on the other hand, also has a low scatter similar to \YT. However, the deviation from a power law relation is much larger than \YT, and the break from power law occurs at a higher halo mass.

\begin{figure}
\centering
\includegraphics[scale=0.5,keepaspectratio=true]{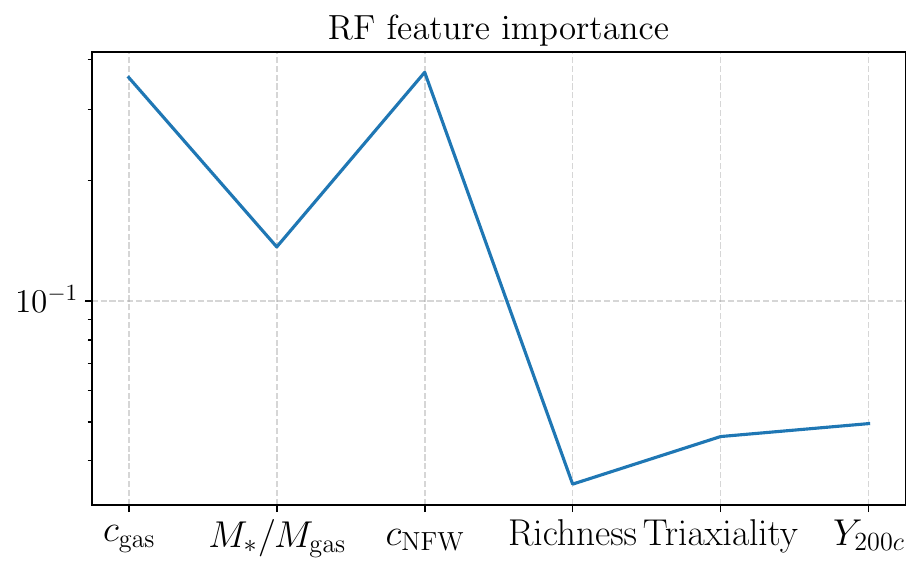}
\caption{The importance of different input variables for the random forest (RF) prediction. The predictions of RF corresponding to the three most important variables is shown in Fig.~\ref{fig:RF_scatter_reduction} of the main text.}
\label{Feature_importance}
\end{figure}

We have primarily used the random forest regressor (RF) as feature selection tool in the main text. We show the relative importance of different input features for the RF prediction in Fig.~\ref{Feature_importance}. We indeed see that $c_\mathrm{gas}, M_*/M_\mathrm{gas}$ and $c_\mathrm{NFW}$ are the most important features.

\begin{figure}
\centering
\includegraphics[scale=0.55,keepaspectratio=true]{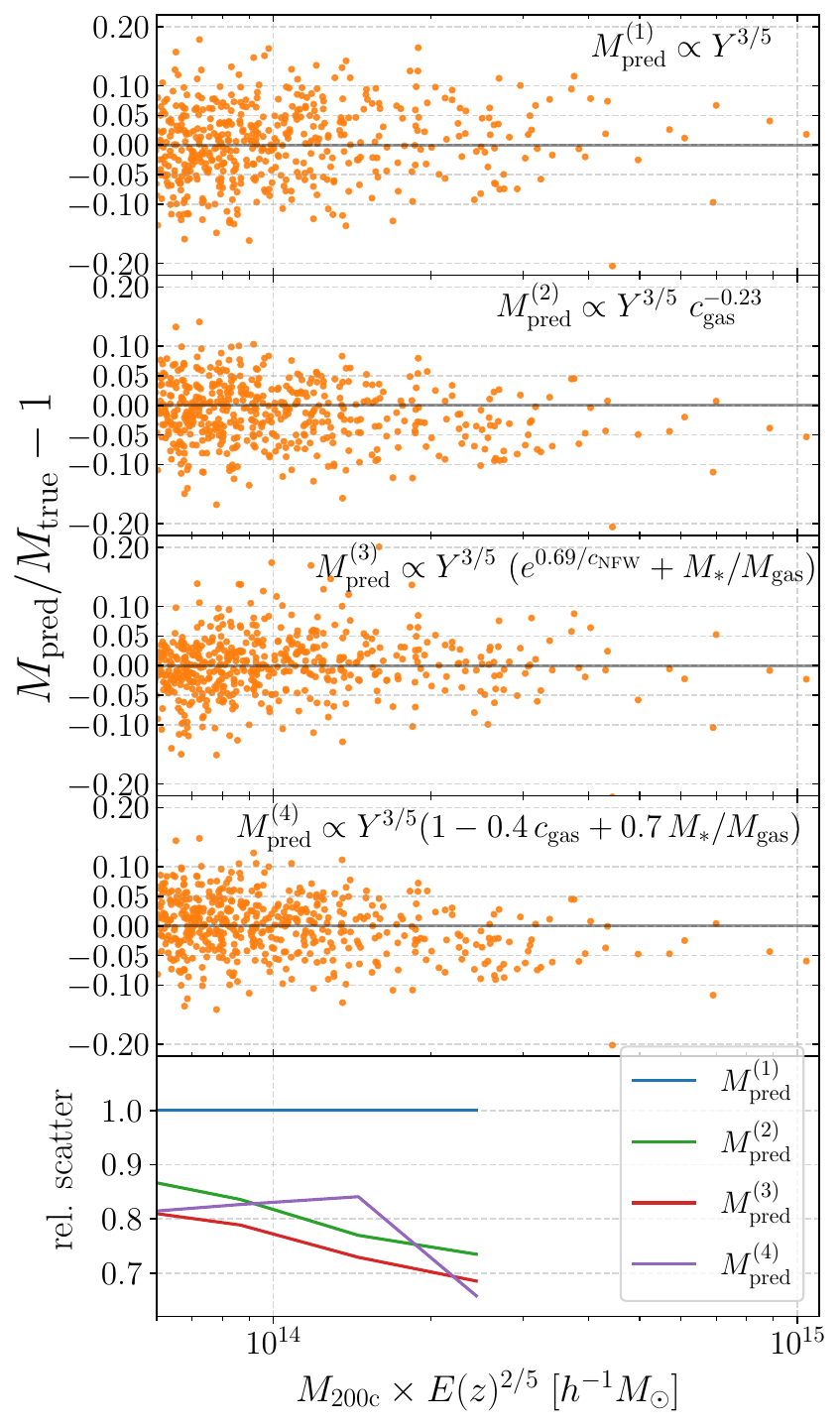}
\caption{Same as Fig.~\ref{fig:SyReg_conc} in the main text, but showing the performance of a few additional equations obtained from the symbolic regressor.
}
\label{fig:SyReg_extra}
\end{figure}

In the final list of formulae obtained from \textsc{PySR}, we choose the simplest ones to compare in Fig.~4 of the main text. We show a few additional results and compare their performance in reducing the scatter in Fig.~\ref{fig:SyReg_extra}.

\begin{figure*}
\centering
\includegraphics[scale=0.45,keepaspectratio=true]{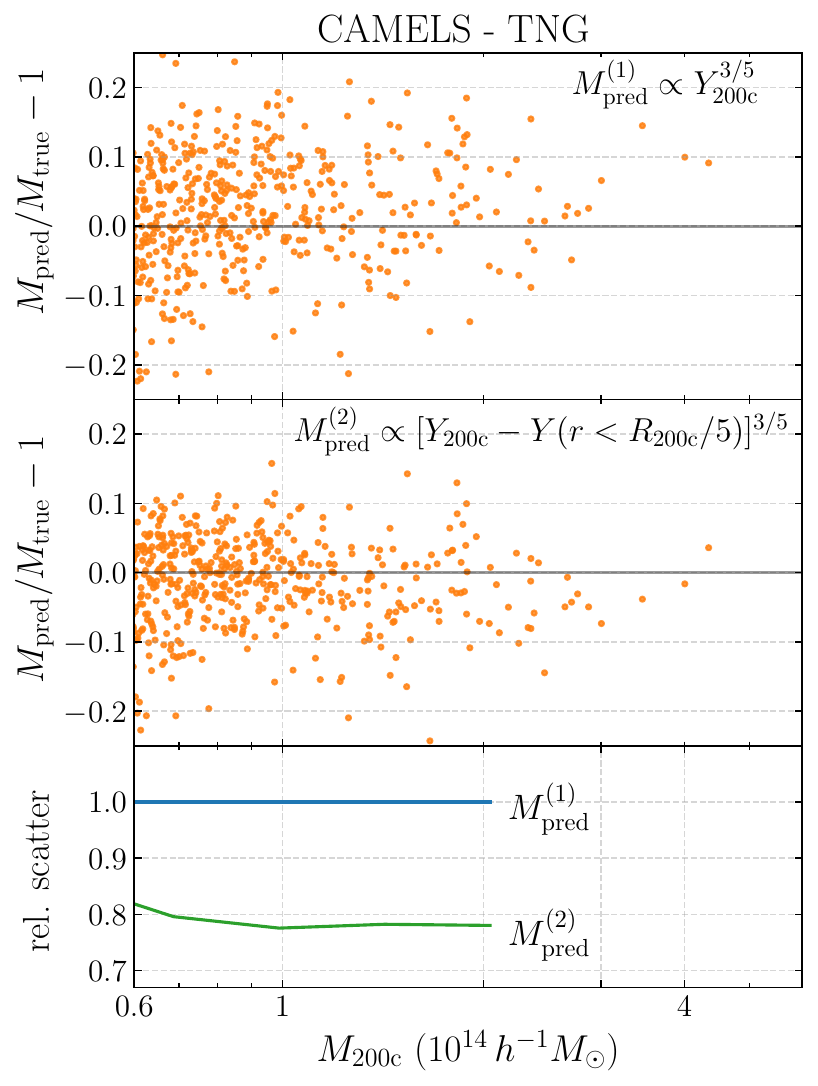}
\includegraphics[scale=0.45,keepaspectratio=true]{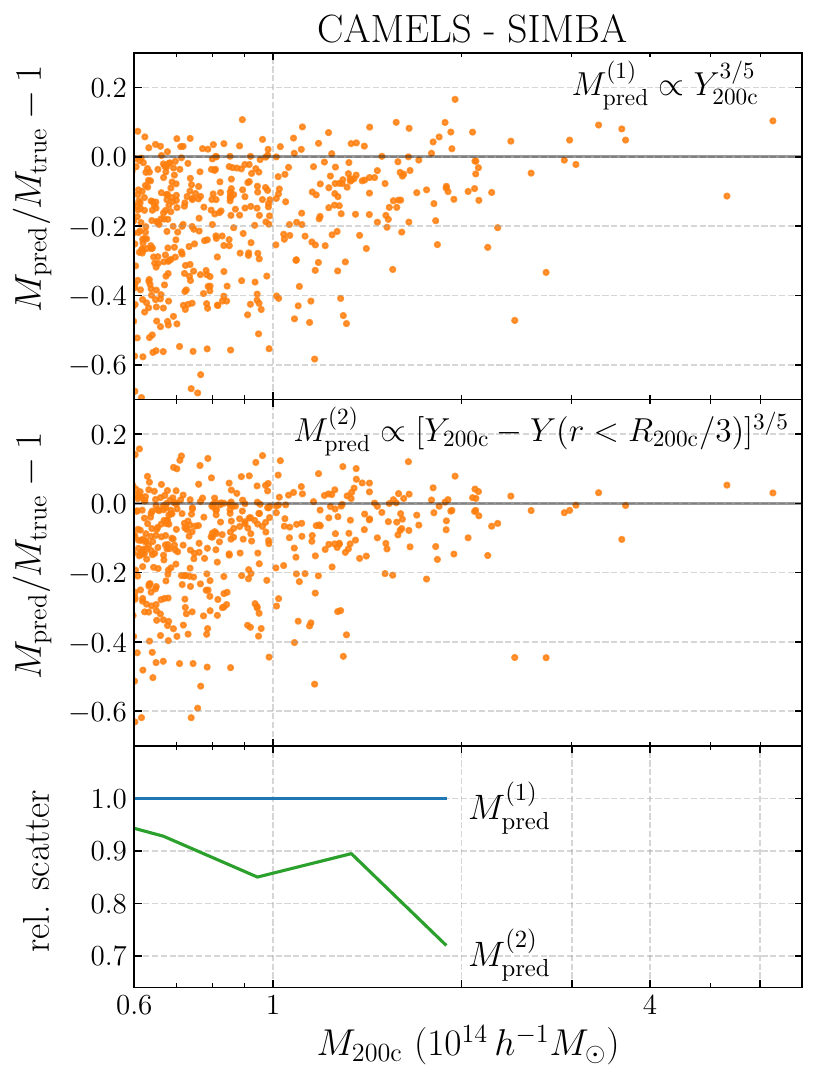}
\caption{Same as Fig.~\ref{fig:CenterExcised} in the main text, but for clusters in the CAMELS simulation suite instead of TNG300. We see a similar reduction in scatter once the cores of these clusters are excised.}
\label{fig:CenterExcised_camels}
\end{figure*}

In Fig.~\ref{fig:CenterExcised} of the main text, we show the reduction in the $Y-M$ scatter when cores of clusters in the TNG300 simulation are excised. We show a similar plot for the CAMELS simulations in Fig.~\ref{fig:CenterExcised_camels}. 

\begin{figure}
\centering
\includegraphics[scale=0.6,keepaspectratio=true]{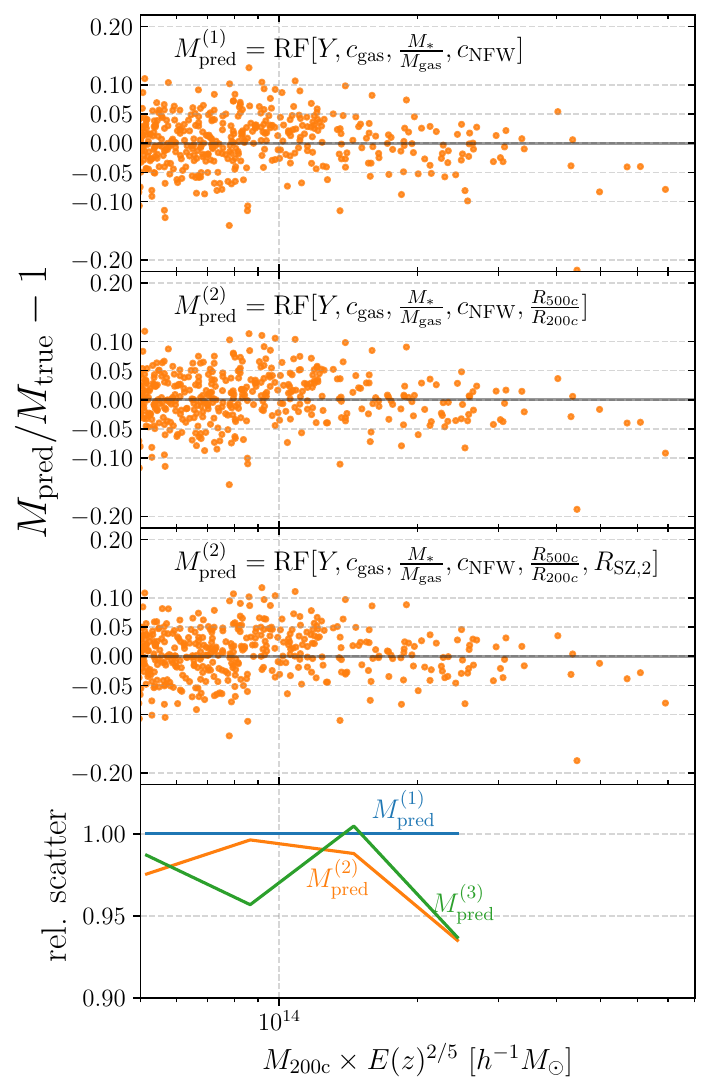}
\caption{Same as Fig.~\ref{fig:RF_scatter_reduction} in the main text but adding to the RF training set two additional parameters which have been proposed in the previous literature to augment the $Y-M$ relation. $R_{500c}/R_{200c}$ corresponds to an analogue of the halo concentration and $R_\mathrm{SZ,2}$ corresponds to an analogue of the SZ half-light radius (see the text for further details).}
\label{RF_prev_literature}
\end{figure}

Finally, in Fig.~\ref{RF_prev_literature}, we show the comparison of results from RF on adding two extra parameters (which were used in the previous literature to augment the $Y-M$ relation). Ref.~\cite{Yan10} proposed using $R_{500c}/R_{200c}$ as an analogue of the NFW concentration. Ref.~\cite{Afs08} proposed using half-light radius for the SZ flux, which is defined as the radius of the sphere that contains half of the total SZ flux. We only collect data for the clusters until 1.5$\times R_{200c}$ and find that the integrated $Y$ has not converged until this radius. We therefore use a different version of the half-light radius than the one proposed in \cite{Afs08}; in our case, $R_\mathrm{SZ,2}$ is obtained from the condition: $Y(r<R_\mathrm{SZ,2}) = Y_{200c}/2$.  


%

\end{document}